\documentstyle{article}
\newcommand{\eff}{\mbox{\footnotesize eff}}
\newcommand{\mf}[1]{\mbox{\footnotesize #1}}
\renewcommand{\(}{\left (}
\renewcommand{\)}{\right )}
\newcommand{\beq}{\begin{equation}}
\newcommand{\eeq}{\end{equation}}
\newcommand{\beqa}{\begin{eqnarray}}
\newcommand{\eeqa}{\end{eqnarray}}
\newcommand{\sij}{s_{ij}}
\newcommand{\vij}{v_{ij}}
\newcommand{\uij}{u_{ij}}
\newcommand{\Pij}{P_{ij}}
\input{psfig}
\begin{document}
\begin{titlepage}
\begin{flushright}
LU TP 96-6\\
\today \\
\end{flushright}
\vspace{0.8in}
\LARGE
\begin{center}
{\bf Airline Crew Scheduling with Potts Neurons}\\
\vspace{.3in}
\large
Martin Lagerholm\footnote{martin@thep.lu.se},
Carsten Peterson\footnote{carsten@thep.lu.se} and
Bo S\"{o}derberg\footnote{bs@thep.lu.se}\\
\vspace{0.1in}
Department of Theoretical Physics, University of Lund\\ S\"{o}lvegatan 14A,
S-223 62 Lund, Sweden\\
\vspace{0.15in}

Submitted to {\it Neural Computation}

\end{center}
\vspace{0.2in}
\normalsize

Abstract:

A Potts feedback neural network approach for finding good solutions to
resource allocation problems with a non-fixed topology is
presented. As a target application the airline crew scheduling problem
is chosen. The topological complication is handled by means of a
propagator defined in terms of Potts neurons. The approach is tested
on artificial random problems tuned to resemble real-world
conditions. Very good results are obtained for a variety of problem
sizes. The computer time demand for the approach only grows like
$\mbox{(number of flights)}^3$. A realistic problem typically is
solved within minutes, partly due to a prior reduction of the problem
size, based on an analysis of the local arrival/departure structure at
the single airports.
%\BS{Something on reduction?}

\end{titlepage}

\newpage

%\large
%\normalsize

\section*{Introduction}

Feedback neural network have in the last decade emerged as a useful
method to obtain good approximate solutions to various resource
allocation problems \cite{tank,pet2,gis1,gis2}. Most applications have
concerned fairly academic problems like the traveling salesman problem
(TSP) and various graph partition problems \cite{tank,pet2}. In refs.
\cite{gis1,gis2} high school scheduling was approached. The typical
approach proceeds in two steps: (1) map the problem onto a neural
network (spin) system with a problem-specific energy function, and (2)
minimize the energy by means of deterministic mean field (MF)
equations, which allow for a probabilistic interpretation. Two basic
mapping variants are common: a hybrid (template) approach \cite{dur},
and a ``purely'' neural one. The template approach is advantageous
e.g. for low-dimensional geometrical problems like the TSP, whereas
for generic resource allocation problems, a purely neural Potts
encoding is preferable.
%
%Scheduling problems occur at different levels in the transportation
%industry. For airlines, one schedules (1) flights based on load
%predictions, (2) aircrafts based on flight time tables and (3) crews
%based on flight time tables.

A very challenging resource allocation problem is airline crew
scheduling, where a given flight schedule is to be covered by a set of
crew {\it rotations}, each consisting in a connected sequence of
flights ({\it legs}), starting and ending at a given home base ({\it
hub}). The total crew waiting time is then to be minimized, subject to a
number of restrictions on the rotations. This application differs
strongly from e.g. high school scheduling \cite{gis1,gis2} in the
existence of non-trivial topological restrictions. A similar structure
occurs in multi-task phone routing.

A common approach to this problem consists in converting it into a
{\it set covering problem}, by (1) generating a large pool of legal
rotation templates, and (2) seeking a subset of the templates that
precisely covers the entire flight schedule. Solutions to the set
covering problem are then found with linear programming techniques or
feedback neural network methods \cite{ohl}. A disadvantage with this
method is that the rotation generation for computational reasons has
to be non-exhaustive for a large problem; thus, only a fraction of the
solution space is available.
%hence the solution space is artificially
%limited.
% in a somewhat arbitrary fashion.

The approach to the crew scheduling problem developed in this letter
%employed in this paper
is quite different, and proceeds in two steps. First, the full
solution space is narrowed down using a reduction technique that
removes a large part of the sub-optimal solutions.
%The suffers
%from no such limitations: solutions are searched for within
%In this letter we develop an approach to airline crew scheduling
Then, a mean field annealing approch based on Potts neurons is
applied, where a novel key ingredient is the use of a propagator
formalism for handling topology, leg-counting, etc.

The method, which is explored on random artificial problems resembling
real-world situations, performs well with respect to quality, with a
computational requirement that grows like $N_f^3$, where $N_f$ is the
number of flights.

\section*{Nature of Problem}

\begin{figure} [t,b]
\hbox{\vbox{
\begin{center}
\mbox{\psfig{figure=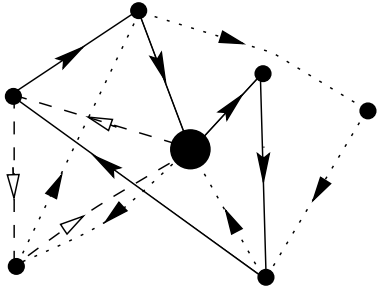,width=5cm}}
\end{center}
}}
 \caption{Schematic view of the 3 crew rotations starting and ending in a hub.}
 \label{fig_ap}
\end{figure}

Typically, a real-world flight schedule has a basic period of one
week. Given such a schedule in terms of a set of $N_f$ weekly
flights, with specified times and airports of departure and arrival, a
crew is to be assigned to each flight such that the total crew
waiting time is minimized, subject to the constraints:
\begin{itemize}
\item Each crew must follow a closed loop -- {\em rotation} --
	starting and ending at the hub (see fig. \ref{fig_ap}).
\item The number of flight legs in a rotation must not exceed a given
	upper bound.
\item The total duration (flight + waiting time) of a rotation is
	similarly bounded.
\end{itemize}
These are the crucial and difficult constraints; in a real-world
problem there are often some 20 additional ones, which we for
simplicity neglect; they constitute no additional challenge from an
algorithmic point of view.

Without the above constraints, the problem would reduce to the {\em
local problem} of minimizing waiting times independently at each
airport; this can be done exactly in polynomial time. It is the global
structural requirements that make the crew scheduling problem a
challenge.

\section*{Reduction of Problem Size -- Airport Fragmentation}

Prior to developing our artificial neural network method, we will
describe a technique to reduce the size of problem, based on the local
flight structure at each airport.

%As mentioned above, upon neglecting the constraints the crew
%scheduling problem factorizes into a set of polynomially solvable
%subproblems, each being local to an airport and consisting in blindly
%minimizing the total local waiting time.
%
With the waiting time between an arriving flight $i$ and a departing
flight $j$ defined as
\beq
	t^{\mf{(w)}}_{ij} =
	\( t^{\mf{(dep)}}_j -
	t^{\mf{(arr)}}_i \) \bmod \mbox{period,}
\eeq
the total waiting time for a given problem can only change by an
integer times the period.
%will be quantized in units of the period.
%
By demanding a minimal waiting time, the local problem (neglecting the
global constraints) at each airport typically can be split up into
independent subproblems, each containing a subset of the arrivals and
an equally large subset of the departures. Some of these are trivial,
forcing the crew of an arrival to continue to a particular departure.
The minimal total wait-time for the local problem is straight-forward
to compute, and will be denoted by $T^{\mf{wait}}_{\mf{min}}$.

Similarly, by demanding a solution (assuming one exists) with
$T^{\mf{wait}}_{\mf{min}}$ also for the constrained global problem,
this can be reduced as follows:
\begin{itemize}
\item {\em Airport fragmentation}: Divide each airport into {\em
	effective airports} corresponding to the non-trivial local
	subproblems.
\item {\em Flight clustering}: Join every forced sequence of flights
	into one effective {\em composite flight}, which will thus
	represent more than one leg and have a formal duration defined
	as the sum of the durations of its legs and the waiting times
	between them.
\end{itemize}
The {\em reduced problem} thus obtained differs from the original
problem only in an essential reduction of the sub-optimal part of the
solution space; the part with minimal waiting time is unaffected by
the reduction. The resulting information gain, taken as the natural
logarithm of the decrease in size of the solution space, empirically
seems to scale approximately like $1.5 \times$ (number of flights),
and ranges from 100 to 2000 for the problems probed.

The reduced problem may in most cases be further separated into a set
of independent {\em subproblems}, that can be solved one by one. Some
of the composite flights will formally arrive at the same effective
airport they started from. This does not pose a problem. Indeed, if
the airport in question is the hub, such a single flight constitutes a
separate (trivial) subproblem, representing an entire forced
rotation. Typically, one of the subproblems will be much larger than
the rest, and will be referred to as the {\em kernel problem}, while
the remaining subproblems will be essentially trivial.

In the formalism below, we allow for the possibility that the problem
% assume that the original problem
to be solved has been reduced as described above, which means that
flights may be composite.
%and that only the kernel problem is dealt with.

\section*{Potts Encoding}

A naive way to encode the crew scheduling problem would be to
introduce Potts spins in analogy with what was done in
refs. \cite{gis1,gis2} where each event (lecture) is mapped onto a
resource unit (lecture-room, time-slot). This would require a Potts
spin for each flight to handle the mapping onto crews.

Since the problem consists in linking together sequences of
(composite) flights such that closed loops are formed, it appears more
natural to choose an encoding where each flight $i$ is mapped, via a
Potts spin, onto the flight $j$ to follow it in the rotation:
\[ \sij =\left\{ \begin{array}{ll}
 $1$ & \mbox{if flight $i$ precedes flight $j$ in a rotation} \\
 $0$ & \mbox{otherwise}
\end{array}
\right. \]
where it is understood that $j$ be restricted to depart from the
(effective) airport where $i$ arrives.
In order to ensure that proper closed loops are formed, each flight
has to be mapped onto precisely one other flight (or terminate a
rotation, in which case it is formally mapped on a dummy flight
available only at the hub). This restriction is inherent in the Potts
spin, defined to have precisely one component ``on'':
\beq
\label{potts}
	\sum_j \sij = 1
\eeq
%

%Due to the global topological constraints, and the variability of the
%number of legs in a rotation, it is not possible (at least with our
%encoding)\BS{It might be possible with other encodings!} to devise a
%polynomial energy function for this system\BS{In fact, we have no
%energy function at all}.
%
%We circumvent this problem (No, we face it like men/BS)
%
Global topological properties, leg-counts and durations of rotations,
etc., cannot be described in a simple way by polynomial functions of
the spins.  Instead, they are conveniently handled by means of a {\em
propagator} matrix, ${\bf P}$, defined a in terms of the Potts spin
matrix ${\bf s}$ by
\beq
\label{prop}
	\Pij = \( ( {\bf 1}-{\bf s} )^{-1} \)_{ij} = \delta
		_{ij} + \sij + \sum_k s_{ik} s_{kj} + \sum_{kl}
		s_{ik}s_{kl}s_{lj} + \sum_{klm} s_{ik}s_{kl}
		s_{lm}s_{mj} + \ldots
\eeq
\begin{figure}[t,b]
%\hbox{\psfig{figure=/usr/users/martin/manus/prop.ps,width=15cm,height=.5cm}}
\hbox{\psfig{figure=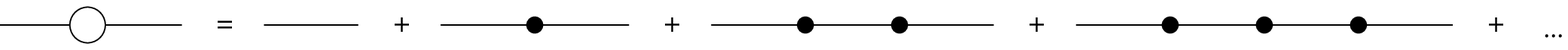,width=14.8cm,height=.4cm}}
\vspace{0.2in}
\caption{Expansion of the propagator $\Pij$ ($\bigcirc$) in terms of
 	$\sij$. A line represents a flight, and ($\bullet$) a
 	touch-down.}
\label{fig_prop}
\end{figure}
A pictorial expansion of the propagator is shown in
fig. \ref{fig_prop}. The interpretation is obvious: $\Pij$ counts the
number of connecting paths from flight $i$ to $j$. Similarly, an
element of the matrix square of $P$,
\beq
\label{P2}
	\sum_k P_{ik} P_{kj} = \delta_{ij} + 2 s_{ij} + 3 \sum_k s_{ik}s_{kj} + \ldots
\eeq
counts the total number of (composite) legs in the connecting paths,
while the number of proper legs is given by
\beq
\label{PLP}
	\sum_k P_{ik} L_k P_{kj} = L_i \delta_{ij}
	+ s_{ij} \( L_i + L_j \)
	+ \sum_k s_{ik} s_{kj} \( L_i + L_k + L_j \)
	+ \ldots
\eeq
where $L_k$ is the intrinsic number of single legs in the composite
flight $k$. Thus, $L_{ij} \equiv \sum_k P_{ik} L_k P_{kj}/\Pij$ gives
the {\em average leg count} of the connecting paths.  Similarly, the
{\em average duration} (flight + waiting time) of the paths from $i$
to $j$ amounts to
\beq
	T_{ij} \equiv \frac{\sum_k P_{ik} t^{\mf{(f)}}_k P_{kj} +
	\sum_{kl} P_{ik} t^{\mf{(w)}}_{kl} s_{kl} P_{lj}}
	{P_{ij}}
%\\ \nonumber
%	& = & t^{\mf{(f)}}_i \delta_{ij}
%	+ s_{ij} \( t^{\mf{(f)}}_i + t^{\mf{(w)}}_{ij} + t^{\mf{(f)}}_j \)
%	+ \sum_k s_{ik} s_{kj} \(
%		t^{\mf{(f)}}_i + t^{\mf{(w)}}_{ik} + t^{\mf{(f)}}_k
%		+ t^{\mf{(w)}}_{kj} + t^{\mf{(f)}}_j
%		\)
%	+ \ldots
\eeq
where $t^{\mf{(f)}}_i$ denotes the duration of the composite flight
$i$, including the embedded waiting time.

Furthermore, any improper loops (such as obtained e.g. if two flights
are mapped onto each other) will make ${\bf P}$ singular -- for a
proper set of rotations, $\det {\bf P} = 1$.

%\section*{Mean Field Approach: Potts Neurons}
\section*{Mean Field Approach}
%Updating Equations

We use a {\em mean field} (MF) annealing approach in the search for
the global minimum. The discrete Potts variables, $s_{ij}$, are
replaced by continuous MF Potts {\em neurons}, $v_{ij}$. They
represent thermal averages $<s_{ij}>_T$, with $T$ an artificial
temperature to be slowly decreased (annealed), and have an obvious
interpretation of probabilities (for flight $i$ to be followed by
$j$). The corresponding probabilistic propagator ${\bf P}$ will be
defined as the matrix inverse of ${\bf 1} - {\bf v}$.

The neurons are updated by iterating the MF equations
\beq
\label{mft}
	\vij = \frac{\exp(u_{ij}/T)}{\sum_k \exp(u_{ik}/T)}
\eeq
for one flight $i$ at a time, by first zeroing the $i$:th row of ${\bf
v}$, and then computing the relevant {\em local fields} $\uij$
entering eq. (\ref{mft}) as
\beq
\label{uij}
	\uij =
	- c_1 t_{ij}^{\mf{(w)}}
	- c_2 P_{ji}
	- c_3 \sum_k v_{kj}
	- c_4 \Psi \( T^{\mf{(ij)}}_{\mf{rot}} - T^{\mf{max}}_{\mf{rot}} \)
	- c_5 \Psi \( L^{\mf{(ij)}}_{\mf{rot}} - L^{\mf{max}}_{\mf{rot}} \)
\eeq
where $j$ is restricted to be a possible continuation flight to
$i$. In the first term, $t_{ij}^{\mf{(w)}}$ is the local waiting time
between flight $i$ and $j$. The second term suppresses improper loops,
and the third term is a soft exclusion term, penalizing solutions where
two flights point to the same next flight. In the fourth and fifth
terms, $L^{\mf{(ij)}}_{\mf{rot}}$ stands for the total leg count and
$T^{\mf{(ij)}}_{\mf{rot}}$ for the duration of the rotation if $i$
where to be mapped onto $j$, and amount to
\begin{eqnarray}
\label{L}
% No -2, since L_a = L_b = 0 /BS
	L^{\mf{(ij)}}_{\mf{rot}} & = & L_{ai} + L_{jb}
%	\frac{\sum_k P_{ak} L_k P_{ki}} {P_{ai}}
%	+ \frac{\sum_k P_{jk} L_k P_{kb}} {P_{jb}}
\\ \label{T}
	T^{\mf{(ij)}}_{\mf{rot}} & = & T_{ai} + t^{\mf{(w)}}_{ij} + T_{jb}
%	\frac{\sum_k P_{ak} t^{\mf{(f)}}_k P_{ki}
%	+ \sum_{kl} P_{ak} t^{\mf{(w)}}_{kl} v_{kl} P_{li}} {P_{ai}}
%	+ t^{\mf{(w)}}_{ij}
%	+ \frac{\sum_k P_{jk} t^{\mf{(f)}}_k P_{kb}
%	+ \sum_{kl} P_{jk} t^{\mf{(w)}}_{kl} v_{kl} P_{lb}} {P_{jb}}
\end{eqnarray}
Here, $a$ and $b$ are auxiliary dummy-flights (of zero duration and
intrinsic leg count) representing the start/end of a rotation -- at
the hub, $a$ is formally mapped onto every departure, and every
arrival is mapped onto $b$.
%
%The denominators $P_{ai}$ and $P_{jb}$ appearing in
%eqs. (\ref{L},\ref{T}) in principle are not necessary; they
%they are included to modulate the dynamical behaviour of the MF
%equations (eqs. (\ref{mft})) such that the different
%$v_{ij}$ settle to $0$ or $1$ respectively at a more or less uniform
%speed as $T$ is decreased.
%
The penalty function $\Psi$, used to enforce the inequality
constraints \cite{ohl}, is defined by $\Psi(x) = x \Theta(x)$ where
$\Theta$ is the Heaviside step function.
Normally, the local fields $u_{ij}$ are derived from a suitable energy
function;
%defined as derivatives (DIFF'S) of an energy function with respect to the
%neurons $v_{ij}$.
however, for reasons of simplicity, some of the terms in eq.
(\ref{uij}) are chosen in a more pragmatic way.
% not to be on this form.

After an initial computation of the propagator ${\bf P}$ from scratch, it
is subsequently updated according to the Sherman-Morrison algorithm
for incremental matrix inversion \cite{numrec}. An update of the $i$:th
row of ${\bf v}$, $v_{ij} \rightarrow v_{ij} + \delta_j$, generates
precisely the following change in the propagator ${\bf P}$:
\beqa
\label{sher1}
	P_{kl} & \rightarrow & P_{kl} + \frac{P_{ki} \, z_{l}}{1-z_{i}}
\\ \label{sher2}
	\mbox{where} \;\;\; z_{l} & = & \sum_{j} \delta_{j} \, P_{jl}
\eeqa
Inverting the matrix from scratch would take $O(N^3)$ operations,
while the (exact) scheme devised above only requires $O(N^2)$ per row.

As the temperature goes to zero, a solution crystallizes in a
winner-takes-all dynamics: for each flight $i$, the largest
$u_{ij}$ determines the continuation flight $j$ to be chosen.

\section*{Test Problems}

In choosing test problems our aim has been to maintain a reasonable
degree of realism, while avoiding unnecessary complication and at the
same time not limiting ourselves to a few real-world problems, where
one can always tune parameters and procedures to get good
performance. In order to accomplish this we have analyzed two typical
real-world template problems obtained from a major airline: one
consisting of long distance (LD), the other of short/medium distance
(SMD) flights.
\begin{figure}
\hbox{
 \psfig{figure=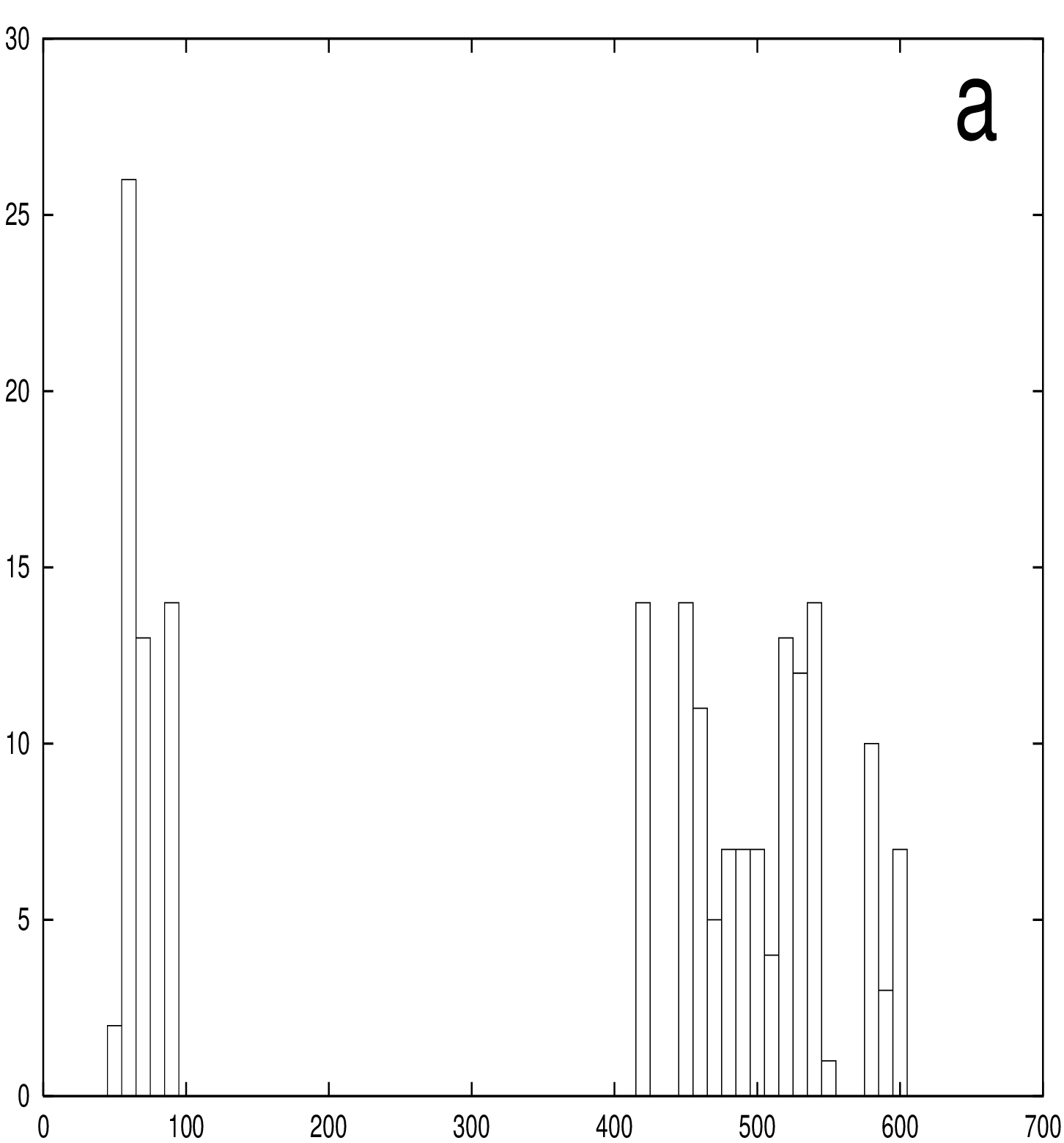,width=7.4cm,height=5cm}
 \psfig{figure=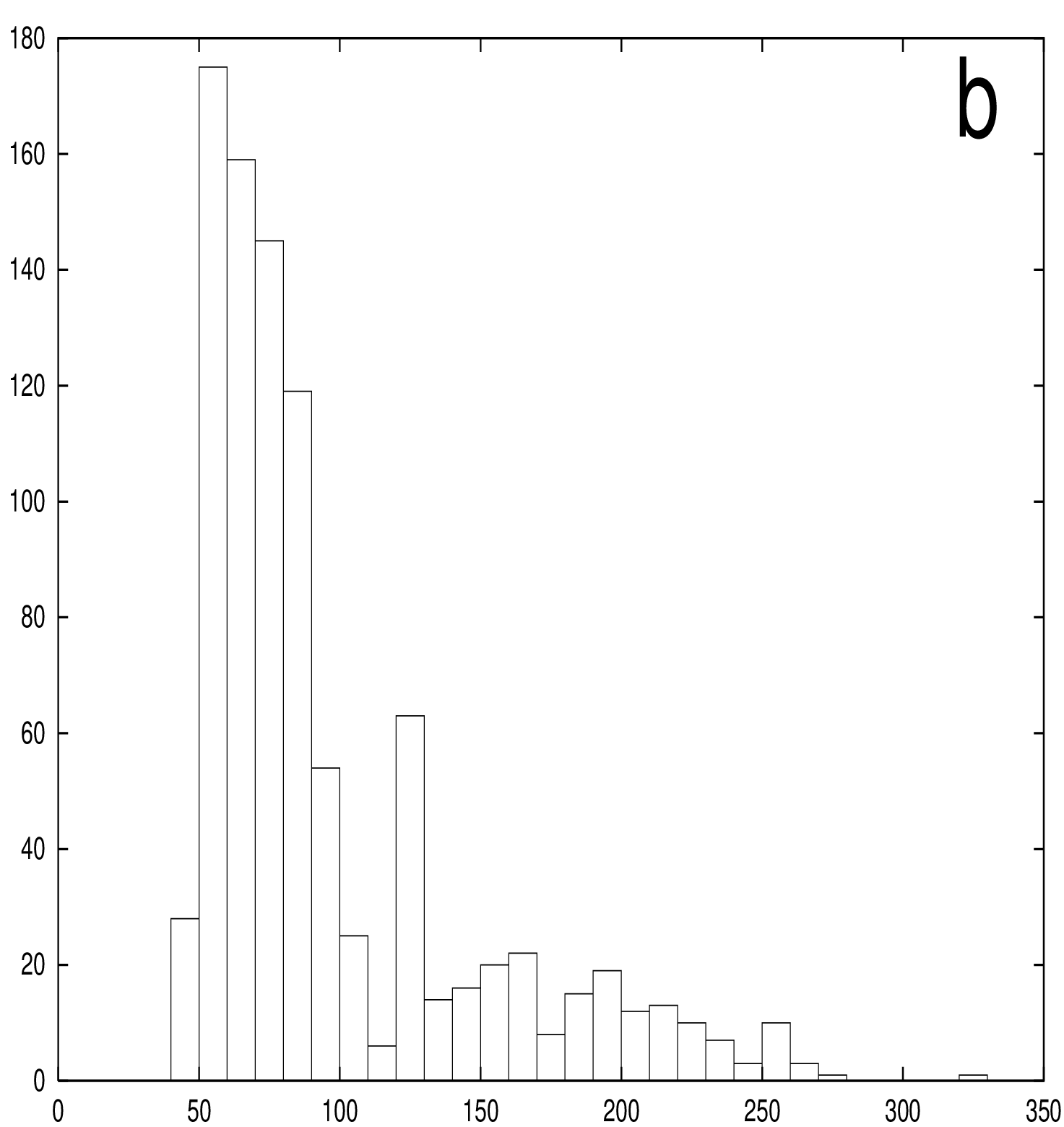,width=7.4cm,height=5cm}
 }
\caption{Fligh time distributions in minutes for {\bf (a)} LD and {\bf
(b)} SMD template problems.}
\label{fig_lh}
\end{figure}
As can be seen from fig. \ref{fig_lh}, LD flight time distributions
are centered around long times, with a small hump for shorter times
representing local continuations of long flights. The SMD flight times
have a more compact distribution.

For each template we have made a distinct problem generator producing
random problems resembling the template. A problem with a specified
number of airports and flights is generated as follows: First, the
distances (flight-times) between airports are chosen randomly from a
suitable distribution. Then, a flight schedule is built up in the form
of legal rotations starting and ending at the hub. For every new leg,
the waiting time and the next airport are randomly chosen in a way
designed to make the resulting problems statistically resemble the
respective template problem.
%so as to resemble a real-world
%schedule.{\ML Reformulate this in some way to make it clearer and
%correct, should also mention that we use different generators for LD
%and SMD problems}

Due to the excessive time consumption
%($O(N!)$)
of the available exact methods, the performance of the Potts approach
cannot be tested against these -- except for in this context
ridiculously small problems, for which the Potts solution quality
matches that of an exact algorithm. For artificial problems of more
realistic size we circumvent this obstacle in the following way: since
problems are generated by producing a legal set of rotations, we add
in the generator a final check that the implied solution yields
$T^{\mf{wait}}_{\mf{min}}$; if not, a new problem is
generated. Theoretically, this might introduce a bias in the problem
ensemble; empirically, however, no problems have had to be
redone. Also the two real problems turn out to be solvable at
$T^{\mf{wait}}_{\mf{min}}$.
%
%\BS{Check this!!} 
%\ML{ Made a check for all (presented) SMD problems; all are generated
%	at the local waiting time!, however 2 had too long loop time and 8
%	had too many legs they were not regenerated, but as you see in
%	the table we solved them anyway. Maybe I should rebuild the
%	generator and rerun those?. Same check for LD problems gives
%	that all gen. problems are solvable at minimal waiting time, in
%	the gen. the max no of legs is built, if too long looptime a
%	rerun is done - this did not happen.}
%

Each problem then is reduced as described above (using a negligible
amount of computer time), and the kernel problem is
% presented to the neural algorithm in the form of
stored as a list of flights, with all traces of the generating
rotations removed.

\section*{Results}

We have tested the performance of the Potts MF approach for both LD
and SMD kernel problems of varying sizes. As an annealing schedule for
the serial updating of the MF eqs. (\ref{mft},\ref{sher1}), we have
used $T_n$=$kT_{n-1}$ with $k=0.9$. In principle, a proper value for
the initial temperature $T_0$ can be estimated from linearizing the
dynamics of the MF equations. We have chosen a more pragmatic
approach: The initial temperature is assigned a tentative value of
$1.0$, which is dynamically adjusted based on the measured rate of
change of the neurons until a proper $T_0$ is found. The values used
for the coefficients $c_i$ in eq. (\ref{uij}) are chosen in an equally
simple and pragmatic way: $c_1 = 1/\mbox{period}$, $c_2 = c_3 = 1$,
while $c_4 = 1/<T^{\mf{rot}}>$ and $c_5 = 1/<L^{\mf{rot}}>$, where
$<T^{\mf{rot}}>$ is the average duration (based on
$T^{\mf{wait}}_{\mf{min}}$) and $<L^{\mf{rot}}>$ the average leg count
per rotation, both of which can be computed beforehand. It is worth
stressing that these parameter settings have been used for the entire
range of problem sizes probed.
%
%\begin{table}
%\begin{center}
%\begin{tabular}{|l|c|c|c|c|c|}
%\hline
% &$c_1 $ &$c_2 $ &$c_3 $ &$c_4 $ &$c_5 $ \\
%\hline \hline
%LD & 1 & 1 & 1 & 1 & 1 \\
%\hline
%SMD & 1 & 1 & 1 & 1 & 1 \\
%\hline
%\end{tabular}
%\end{center}
%\caption{Parameter settings for the penalty terms in eq. (\protect\ref{uij}).}
%\label{fig_c}
%\end{table}
%

When evaluating a solution obtained with the Potts approach, a check
is done as to whether it is legal (if not, a simple post-processor
restores legality -- this is only occasionally needed), then the
solution quality is probed by measuring the excess waiting time $R$,
\beq
\label{perf}
	R = \frac{T^{\mf{wait}}_{\mf{Potts}}-T^{\mf{wait}}_{\mf{min}}}
	{\mbox{period}}, \eeq
which is a non-negative integer for a legal solution.
%; $T^{\mf{wait}}_{\mf{min}}$ denotes the locally minimal waiting time.

For a given problem size, as given by the desired number of airports
$N_a$ and flights $N_f$, a set of 10 distinct problems is generated.
Each problem is subsequently reduced, and the Potts algorithm is
applied to the resulting kernel problem.
%and effective composite flights $N^{\eff}_f$ and
%effective airports $N^{\eff}_a$ are formed.
The solutions are evaluated, and the average $R$ for the set is
computed.
The results for a set of problem sizes ranging from $N_f\simeq 75$ to
$1000$
%
%(corresponding to $N^{\eff}_f=20$ to $N^{\eff}_f=620$
%for real SMD problems)
are shown in tables \ref{fig_LD} and \ref{fig_SMD}, for the two real
problems see table \ref{fig_ori}.
\begin{table}[t]
\begin{center}
\begin{tabular}{|c|c|c|c|c|c|}
\hline
$N _f$ & $N_a$ & $< N^{\eff}_f>$ & $< N^{\eff}_a>$ & $<$ R $>$ & $< $ CPU time $ >$\\
\hline
 75 &  5 &  23 &  8 & 0.0 & 0.0 sec  \\
100 &  5 &  50 & 17 & 0.0 & 0.2 sec  \\
150 & 10 &  55 & 17 & 0.0 & 0.3 sec  \\
200 & 10 &  99 & 29 & 0.0 & 1.3 sec  \\
225 & 15 &  84 & 26 & 0.0 & 0.7 sec  \\
300 & 15 & 154 & 46 & 0.0 & 3.4 sec  \\
\hline
\end{tabular}
\end{center}
\caption{Average performance of the Potts algorithm for {\bf LD}
	problems. The superscript ``eff'' refers to the kernel
	problem, subscript ``$f$'' and ``$a$'' refers to flight
	respectively airport. The averages are taken with 10 different
	problems for each $N_f$. The performance is measured as the
	difference between the waiting time in the Potts and the local
	solutions divided by the period. The CPU time refers to DEC
	Alpha 2000.}
\label{fig_LD}
\end{table}
\begin{table}[h]
\begin{center}
\begin{tabular}{|c|c|c|c|c|c|}
\hline
$N _f$ & $N_a$ & $< N^{\eff}_f>$ & $< N^{\eff}_a>$ & $<$ R $>$ & $< $ CPU time $ >$\\
\hline
%525 & 35  & 213 &  52 & 0.0 & 9.3 sec  \\
600 & 40  & 280 &  64 & 0.0 &  19 sec  \\
675 & 45  & 327 &  72 & 0.0 &  35 sec  \\
700 & 35  & 370 &  83 & 0.0 &  56 sec  \\
750 & 50  & 414 &  87 & 0.0 &  90 sec  \\
800 & 40  & 441 &  91 & 0.0 & 164 sec  \\
900 & 45  & 535 & 101 & 0.0 & 390 sec  \\
1000& 50  & 614 & 109 & 0.0 & 656 sec  \\
\hline
\end{tabular}
\end{center}
\caption{Average performance of the Potts algorithm for {\bf SMD}
	problems. The averages are taken with 10 different problems
	for each $N_f$. Same notation as in table
	\protect\ref{fig_LD}.}
\label{fig_SMD}
\end{table}
\begin{table}[t]
\begin{center}
\begin{tabular}{|c|c|c|c|c|c|c|}
\hline
$N _f$ & $N_a$ & $< N^{\eff}_f>$ & $< N^{\eff}_a>$ & $<$ R $>$ & $< $ CPU time $ >$ & type\\
\hline
189 & 15 &  71 & 24 & 0.0 & 0.6 sec &  LD \\
948 & 64 & 383 & 98 & 0.0 & 184 sec & SMD \\
\hline
\end{tabular}
\end{center}
\caption{Average performance of the Potts algorithm for 10 runs on the
	two {\bf real} problems. Same notation as in table
	\protect\ref{fig_LD}.}
\label{fig_ori}
\end{table}

The results are quite impressive -- the Potts
algorithm has solved all problems, and with a very modest CPU time
consumption,
%
%provide good solution qualities; the CPU time consumption is also very
%modest.
%
of which the major part goes into updating the $P$ matrix. The sweep
time scales like $(N^{\eff}_f)^3 \propto N_f^3$, with a small
prefactor, due to the fast method used, eqs. (\ref{sher1},
\ref{sher2}). This should be multiplied by the number of sweeps needed
-- empirically between 30 and 40, independently of problem
size\footnote{The minor apparent deviation from the expected scaling
in tables \protect\ref{fig_LD}, \protect\ref{fig_SMD} and
\protect\ref{fig_ori} are due to an anomalous scaling of the Digital
DXML library routines employed; the number of elementary operations
does scale like $N_f^3$.}.

\section*{Summary}

We have developed a mean field Potts approach for solving resource
allocation problems with a non-trivial topology. The method is applied
to airline crew scheduling problems resembling real-world situations.

A novel key feature is the handling of global entities, sensitive
to the dynamically changing ``fuzzy'' topology, by means of a propagator
formalism.
Another important ingredient is the problem size reduction achieved by
airport fragmentation and flight clustering, narrowing down the
solution space by removing much of the sub-optimal part.

High quality solutions are consistently found throughout a range of
problem sizes without having to fine-tune the parameters, with a time
consumption scaling as the cube of the problem size. The basic
approach should be easy to adapt to other applications, like
e.g. communication routing.

%\subsection*{Acknowledgements:}
%
%This work was in part supported by the Swedish Board for Industrial
%and Technical Development.

\end{document}